\documentclass[conference, 10pt]{IEEEtran}
\ifCLASSINFOpdf
\else
\fi
\usepackage{graphicx}
\usepackage{color}
\usepackage{placeins}
\usepackage{float}
\usepackage{hyperref}
\usepackage{tabularx,colortbl}
\usepackage{amsmath}
\usepackage{subcaption}
\newcommand{\pvec}[1]{\vec{#1}\mkern2mu\vphantom{#1}}

\begin{document}
%
\title{Extreme Beam-forming with Metagrating-assisted Planar Antennas}

\author{\IEEEauthorblockN{Gengyu Xu, Sean V. Hum and George V. Eleftheriades}
\IEEEauthorblockA{The Edward S. Rogers Sr. Department of
Electrical \& Computer Engineering\\
University of Toronto\\
Toronto, Ontario, Canada\\}}


%


\maketitle

\begin{abstract}
We present a highly efficient metagrating-assisted antenna (MGA) architecture with a simple integrated feed. Power from a localized line source is distributed throughout the arbitrarily large antenna aperture with the help of a passive and lossless electromagnetic metagrating (MG). With appropriately designed meta-wire loading, the omnidirectional source field can be efficiently transformed into directive radiation. To aid the design process, a 2-dimensional volume-surface integral equation framework which accurately predicts the radiation pattern of the MGA is developed. Through constrained optimization, the directivity of the MGA in the desired direction is maximized. In this way, extreme-angle beam steering is demonstrated.
\end{abstract}


%
\IEEEpeerreviewmaketitle

\section{Introduction}
\label{sec:intro}
Electromagnetic metagratings (MGs), which consist of sparse arrays of polarizable ``meta-wires", have been explored as an efficient platform for realizing diffractive wave-transformation devices~\cite{PCBMG}. For certain applications, their simplicity can make them more desirable than conventional gradient metasurfaces (MTSs)~\cite{DBMG}, which are composed of dense lattices of subwavelength meta-atoms. 

In this work, we leverage the diffraction engineering capabilities afforded by MGs to design a low-profile planar antenna for 1D beam-forming. A schematic for the proposed three-layer metagrating-assisted antenna (MGA) is shown in Fig.~\ref{fig:schematic_model}(a). An MG consisting of printed meta-wires resides on the top layer, while a line source is embedded in the middle layer to excite the MG. The bottom layer is a ground plane.

To analyze and design the MGA, we develop a volume-surface integral equation framework which is highly suitable for modelling non-homogenizable diffractive devices. With the proposed design method, we forego the conventionally used local periodic assumption, and explicitly model the mutual coupling between adjacent meta-wires. Furthermore, we can rigorously account for complications associated with realistic devices such as edge diffraction and compact illuminations.

\section{Theory}
\begin{figure}
  \includegraphics[width=\linewidth]{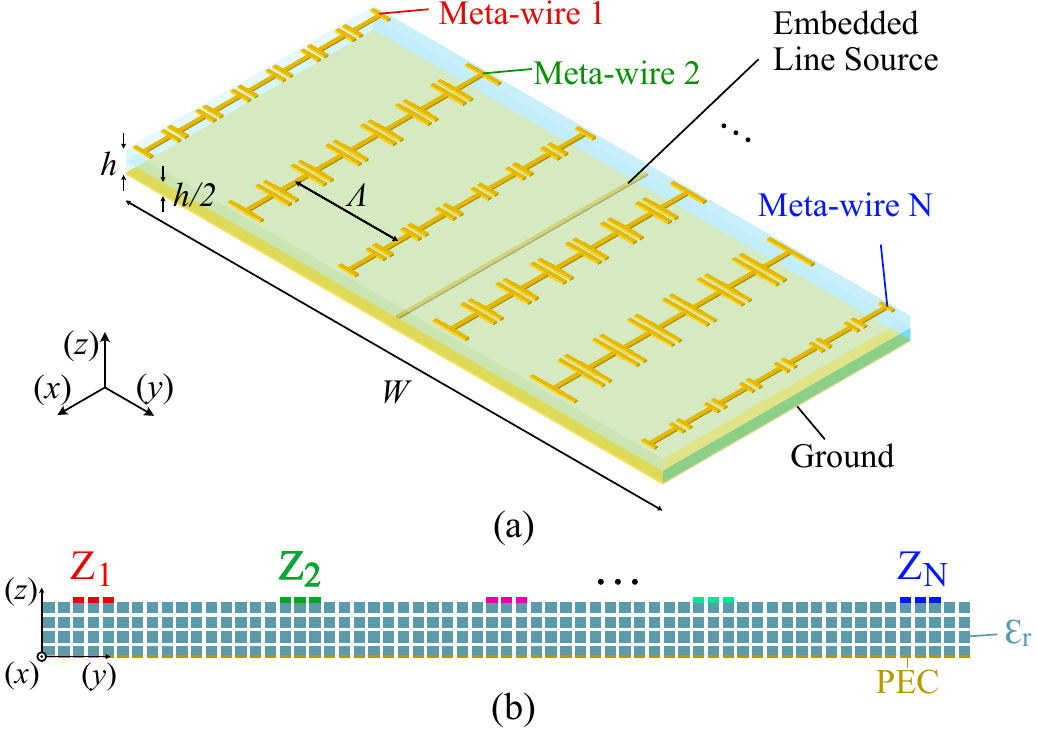}
  \caption{(a) Schematic and (b) theoretical model for the MGA.}
  \label{fig:schematic_model}
\end{figure}

To efficiently model and design the MGA, we consider a simplified 2D representation shown in Fig.~\ref{fig:schematic_model}(b), with the assumption of an $x$-invariant and $x$-polarized (transverse-electric) electric field. We model the $N$ meta-wires with narrow homogeneous strips whose surface electric impedances $Z_n~[\Omega/\mathrm{sq}]$ ($n\in[1,N]$) correspond to the effective distributed meta-wire loads. To avoid the need for controlled power gain and/or loss which can lead to implementation challenges, we assume that the meta-wires are purely reactive ($Z_n=jX_n$). For now, the electric field from the line source ($E_i$) is modelled by a cylindrical wave with frequency $\omega=2\pi f$. We discretize the wires into~$N_w$ segments, and the ground plane into~$N_g$ segments. The dielectric is divided into~$N_v$ rectangular cells.

Under illumination by $E^i$, the meta-wires and the ground plane will carry induced conduction surface current densities $J_w$ and $J_g$ respectively, and produce scattered fields
\begin{equation}
\label{eqn:SIE}
E^s_{\iota}\left(\pvec{\rho}\right) = -\sum_{n=1}^{N_\iota}~\int\limits_{\mathrm{seg}~n}\frac{k\eta}{4}H^{(2)}_0\left(k\left|\vec{\rho}-\pvec{\rho}'\right|\right)J_\iota(\pvec{\rho}')dy',
\end{equation}
with $\iota\in\left\{w,g\right\}$,  $\pvec{\rho}=\hat{y}y+\hat{z}z$ and $\pvec{\rho}'=\hat{y}y'+\hat{z}z'$.
The dielectric substrate will carry a polarization volume current density $J_v$ which has the following contribution to the scattered field:

\begin{equation}
\label{eqn:VIE}
E^s_{v}\left(\pvec{\rho}\right) = -\sum_{n=1}^{N_v}~\iint\limits_{\mathrm{cell}~n}\frac{k\eta}{4}H^{(2)}_0\left(k\left|\vec{\rho}-\pvec{\rho}'\right|\right)J_v(\pvec{\rho}')dy'dz'.
\end{equation}

The total (incident plus scattered) electric field must satisfy

\begin{equation}
\label{eqn:constraints}
E^s_w+E^s_g+E^s_v +E^i= \begin{cases} 0 &\mbox{on ground plane }\\
jX_nJ_w & \mbox{on meta-wire $n$}\\
\frac{J_v}{j\omega\left(\epsilon-\epsilon_o\right)} & \mbox{in dielectric}\end{cases}.
\end{equation}

The system of equations formed by (\ref{eqn:SIE}), (\ref{eqn:VIE}) and (\ref{eqn:constraints}) can be solved, using the method of moments (by point-matching with pulse basis functions), for the unknown current densities. The total fields everywhere can be subsequently evaluated. Therefore, we can establish a closed-form relationship between the far-field radiation pattern ($E^{ff}$) and the  meta-wire loads, the latter of which can be written into a vector $\bar{X}$. Then, a variety of approaches can be taken to shape the radiation pattern of the MGA~\cite{inversion}. As a proof of concept, we maximize the 2D directivity of the MGA in the desired direction~$\theta_o$ using the sequential quadratic programming method:

\begin{equation}
\label{eqn:optimization_problem}
\begin{aligned}
& \underset{\bar{X}}{\text{maximize}}
& &  D_{2D}=\frac{\left|E^{ff}(\theta_o,\bar{X})\right|^2}{\int\limits_0^\pi \left|E^{ff}(\theta,\bar{X})\right|^2 d\theta} \\
& \text{subject to}
& & X_{min}\leq\bar{X}[i]\leq X_{max}, \; i = 1, \ldots, m.
\end{aligned}
\end{equation}
The constraint here limits the effective impedances to values that are practically obtainable using printed meta-wires.

Compared to similar design methods which have previously been used to design reflectarray antennas~\cite{IE}, the proposed method rigoriously account for the dielectric substrate which has a non-negligible impact  on the performance of printed antennas.
 
\section{Results and Discussions}
For illustrative purpose, we design and investigate an MGA operating at 10~GHz, with aperture size $W=7\lambda$ and meta-wire spacing $\Lambda=\lambda/4$. The substrate has a thickness of 0.0085$\lambda$ and relative permittivity $\epsilon_r=3$. We verify the performance of the antenna with finite element simulations in Ansys HFSS, in which the meta-wires are implemented using impedance boundary conditions. The source is modelled as an electric line current. A sub-wavelength slice of the $x$-invariant MGA is placed inside parallel plate waveguides with radiation boundary terminations in order to emulate a 2D environment.

\begin{figure}[t]
     \centering
         \centering
         \includegraphics[width=\linewidth]{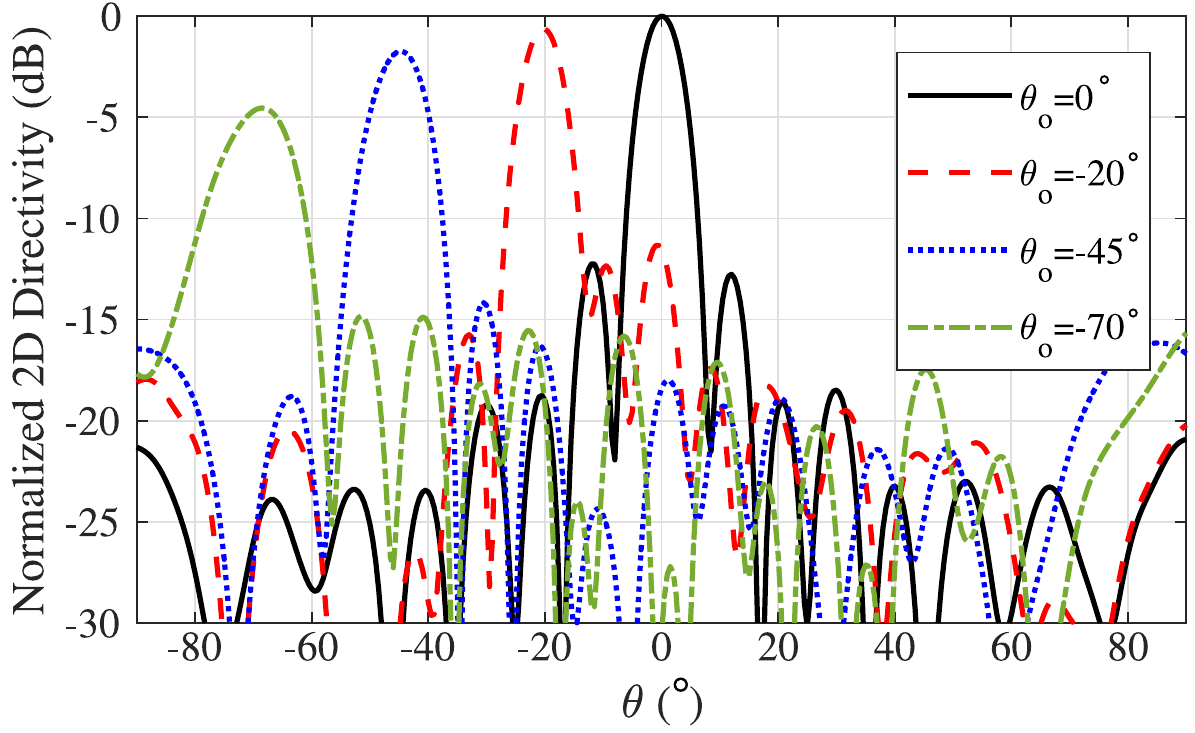}
         \caption{Simulated normalized 2D directivity of the MGA.}
         \label{fig:FF_plot}
\end{figure}

We optimize the meta-wire impedances of the MGA following (\ref{eqn:optimization_problem}) for several different values of $\theta_o$. The resultant 2D directivities for the considered cases (normalized against the maximum directivity for $\theta_o=0^\circ$) are plotted in Fig.~\ref{fig:FF_plot}. Contrary to an infinitely periodic MG which can only produce beams at discrete angles corresponding to the supported Floquet harmonics~\cite{PCBMG}, the proposed finite and  aperiodic device has a continuous scan range. The increased versatility comes at the cost of performance-deteriorating non-idealities such as spurious diffraction, which are especially severe with large scan angles. However, as demonstrated by Fig.~\ref{fig:FF_plot}, we can still achieve a wide scan range of $\pm70^\circ$  while maintaining a very low -10~dB maximum sidelobe level.

\begin{figure}[t]
     \centering
     \begin{subfigure}[b]{0.48\linewidth}
         \centering
         \includegraphics[width=\textwidth]{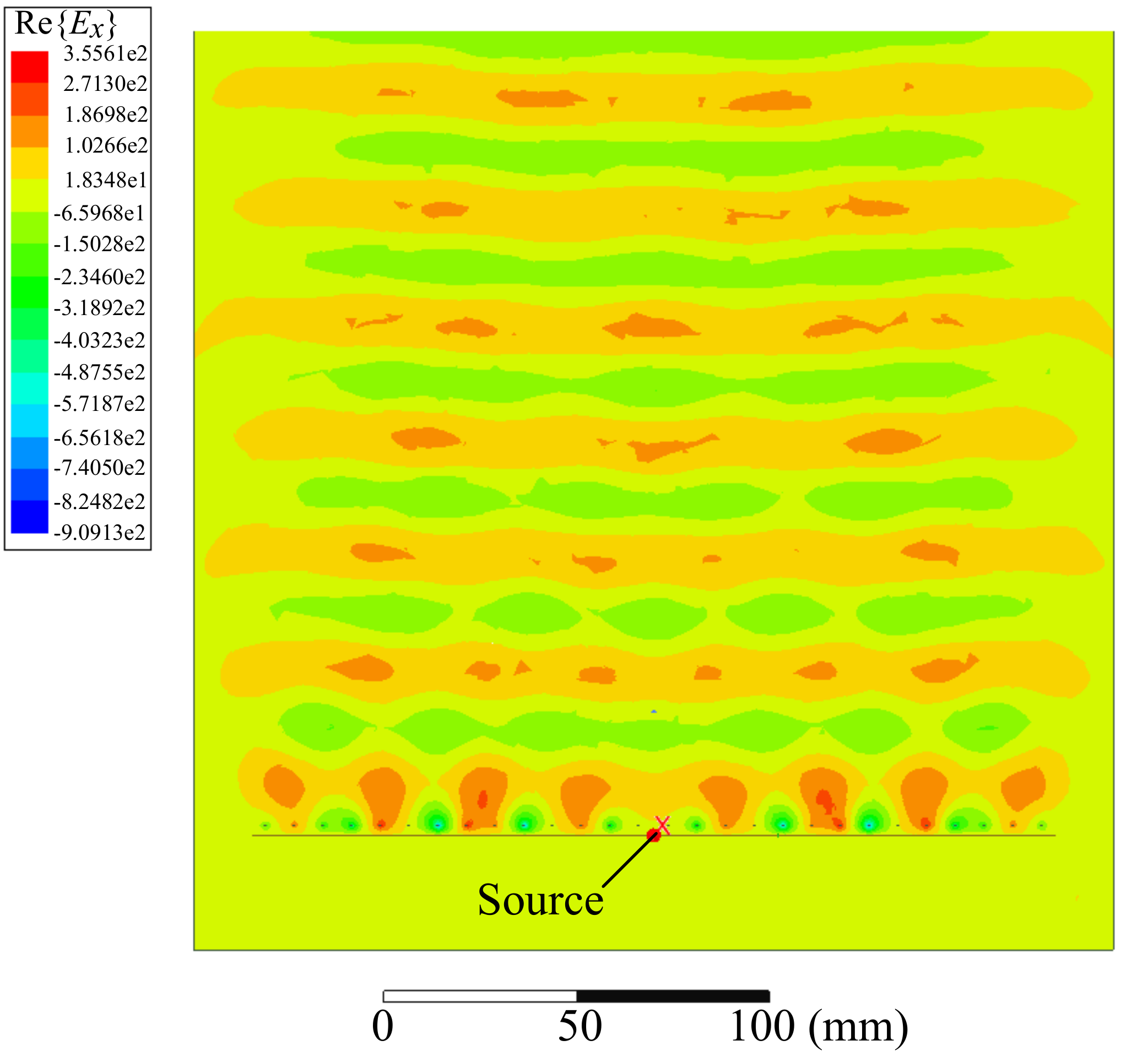}
         \caption{}
         \label{fig:design_a_nf}
     \end{subfigure}
     \hfill
     \begin{subfigure}[b]{0.48\linewidth}
         \centering
         \includegraphics[width=\textwidth]{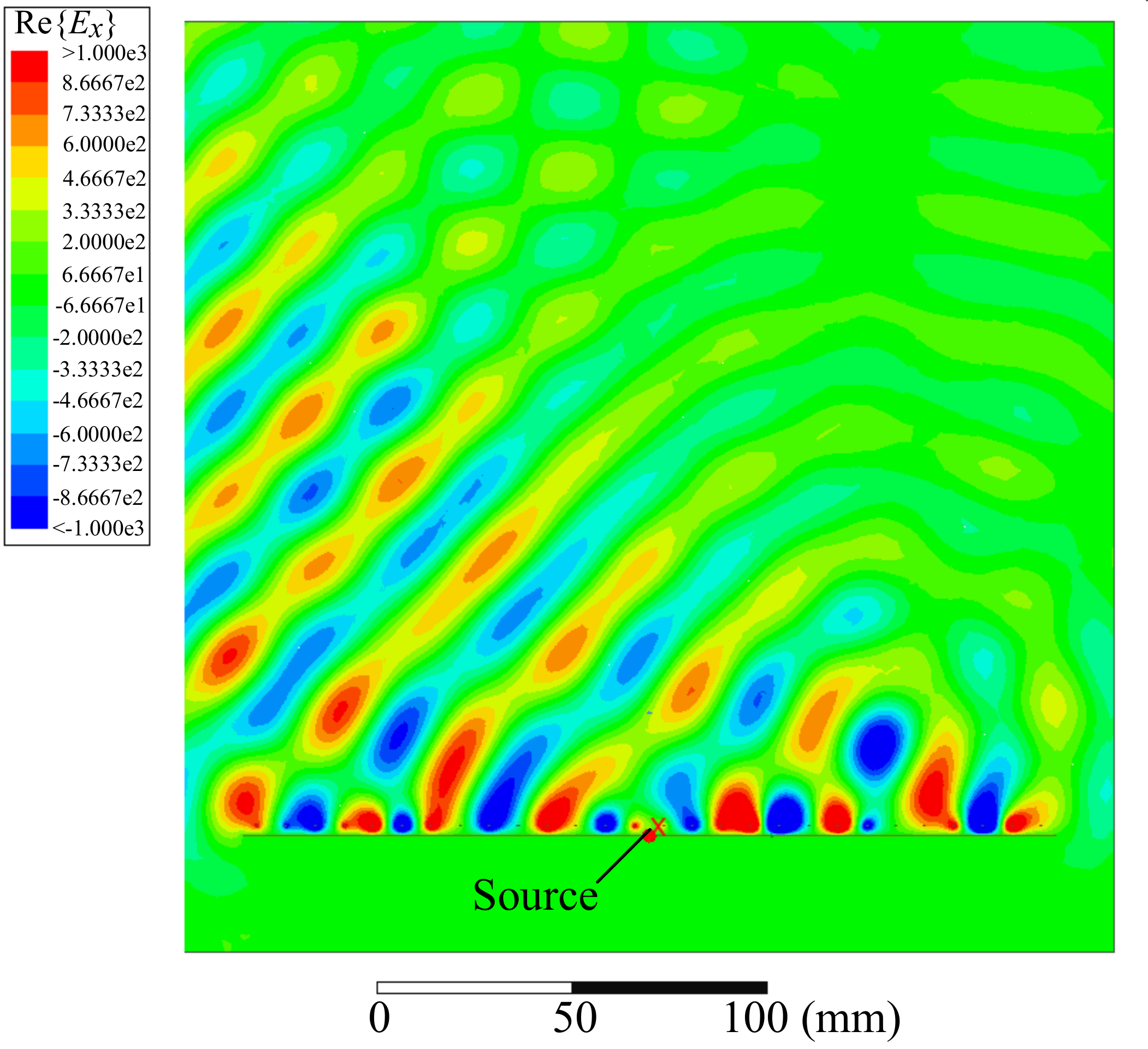}
         \caption{}
         \label{fig:design_a_ff}
     \end{subfigure}
        \caption{Simulated near-field electric field distribution Re$\{E_x\}$ for \mbox{(a) $\theta_o=0^\circ$} and \mbox{(b) $\theta_o = 45^\circ$}.}
        \label{fig:NF_plot}
\end{figure}

To gain a better understanding of the working principle of the MGA, we plot the near-field electric field distribution (Re$\{E_x\}$) for $\theta_o=0^\circ$ and $\theta_o=45^\circ$ in Fig.~\ref{fig:NF_plot}(a) and (b) respectively. It can be seen that the MGA exhibits excellent aperture illumination despite the highly localized line source, regardless of the scan angle. This can be attributed to the reactive MG, which supports and modulates a transverse-electric leaky surface wave that distributes the source power throughout the arbitrarily large aperture, while radiating with the most optimal non-uniform power leakage. Unlike previous works on MTS antennas, we eschew the explicit engineering of the aperture fields in favor of a less informative, but dramatically simpler optimization-based approach. Thus, we circumvent the need to meticulously craft the local phase and leakage constants. Furthermore, since the mutual coupling between all components of the MGA is explicitly modelled, performance degradation due to edge reflections and diffractions, as well as the violation of the local periodic assumption, is inherently minimized.

\end{document}